\documentclass[runningheads]{llncs}
\usepackage[T1]{fontenc}
\usepackage{graphicx}
\usepackage{url} 
\usepackage{booktabs}
\usepackage{amsmath}
\usepackage{bbding}
\begin{document}
%

\title{Zero- and One-Shot Data Augmentation for Sentence-Level Dysarthric Speech Recognition in Constrained Scenarios}

%
\titlerunning{Zero/One-Shot DDA for Sentence-Level DSR in Constrained Scenarios}

\author{Shiyao Wang\orcidID{0009-0006-5786-7109} \and
Shiwan Zhao\orcidID{0000-0001-5068-025X} \and
Jiaming Zhou\orcidID{0009-0002-481
9-4572} \and 
Yong Qin\orcidID{0009-0000-2748-3020}\inst{*}\Envelope
}
\renewcommand{\thefootnote}{}
\footnotetext{* Corresponding author.}

%
\authorrunning{S. Wang et al.}

%
\institute{Nankai University, China \\
\email{wangshiyao@mail.nankai.edu.cn} \\
\email{zhaosw@gmail.com} \\
\email{zhoujiaming@mail.nankai.edu.cn} \\
\email{qinyong@nankai.edu.cn}
}

\maketitle              
\begin{abstract}
Dysarthric speech recognition (DSR) research has witnessed remarkable progress in recent years, evolving from the basic understanding of individual words to the intricate comprehension of sentence-level expressions, all driven by the pressing communication needs of individuals with dysarthria. Nevertheless, the scarcity of available data remains a substantial hurdle, posing a significant challenge to the development of effective sentence-level DSR systems. In response to this issue, dysarthric data augmentation (DDA) has emerged as a highly promising approach. Generative models are frequently employed to generate training data for automatic speech recognition tasks. However, their effectiveness hinges on the ability of the synthesized data to accurately represent the target domain. The wide-ranging variability in pronunciation among dysarthric speakers makes it extremely difficult for models trained on data from existing speakers to produce useful augmented data, especially in zero-shot or one-shot learning settings. To address this limitation, we put forward a novel text-coverage strategy specifically designed for text-matching data synthesis. This innovative strategy allows for efficient zero/one-shot DDA, leading to substantial enhancements in the performance of DSR when dealing with unseen dysarthric speakers. Such improvements are of great significance in practical applications, including dysarthria rehabilitation programs and day-to-day common-sentence communication scenarios. 


\end{abstract}
\section{Introduction}
\label{sec:intro}
\begin{figure}[t]
\centerline{\includegraphics[width=12cm]{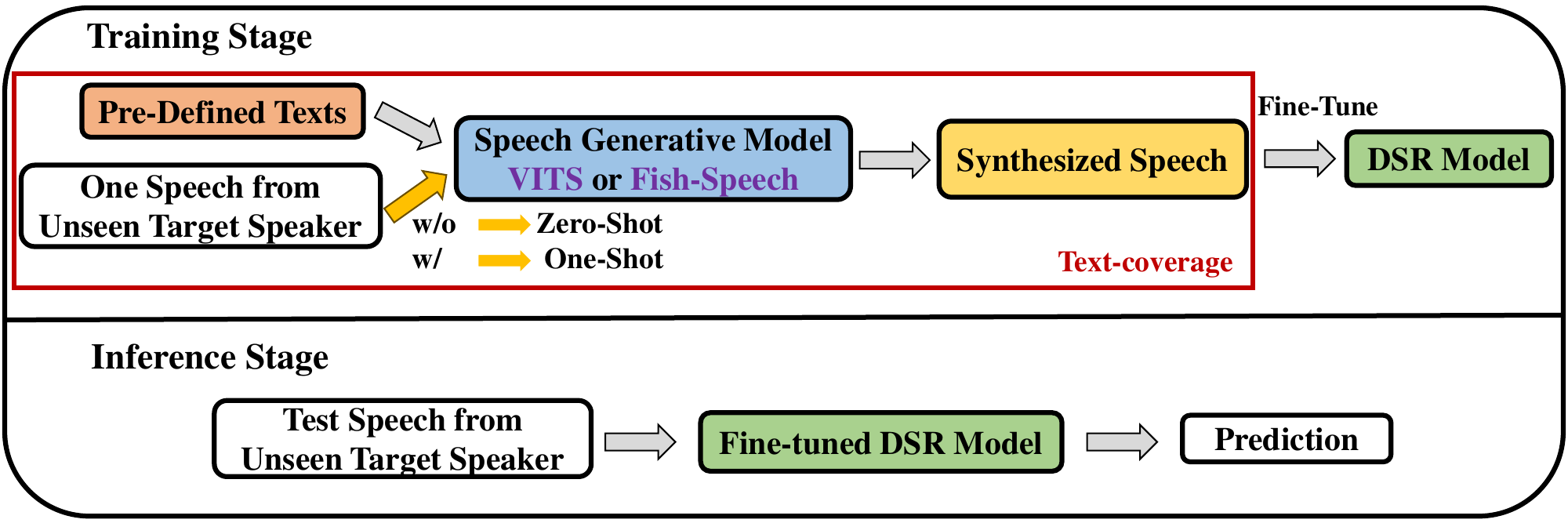}}
\caption{The overview of our Zero-/One-shot DDA pipeline. Training stage: DSR model fine-tuned with \textbf{text-coverage }data. Inference stage: Fine-tuned DSR model used for prediction.}
\label{fig:overview}
\end{figure}
Dysarthria, a speech disorder characterized by impaired vocal control, is commonly associated with neurological conditions such as cerebral palsy and Parkinson's disease. It leads to speech that exhibits pronunciation errors and irregular prosody, significantly hindering automatic speech recognition (ASR) systems trained on typical speech \cite{asr2}. Research in dysarthric speech recognition (DSR) primarily focuses on word recognition \cite{uaspeech,2023dsrsurvey}. Current methodologies include speaker-independent models \cite{si2_dda20223,si4}, which often struggle to generalize to unseen dysarthric speakers due to pronunciation variability, and speaker-dependent models \cite{var}, which necessitate continuous data collection and costly fine-tuning to maintain optimal performance. Wang et al. \cite{pbdsr} proposed a prototype-based, fine-tuning-free classification method for word-level DSR that utilizes speaker-specific pronunciation patterns for rapid and effective adaptation. However, DSR now faces the challenge of achieving sentence-level understanding. Sentence-level DSR is particularly sensitive to pronunciation patterns, including breathing sounds, abnormal pauses (which can lead to insertion errors), and mispronunciations (resulting in replacement errors). Additionally, sentence-level DSR encounters challenges related to larger vocabularies and coarticulation effects, requiring more targeted data for effective adaptation.

Acquiring data from dysarthric speakers presents challenges due to participant limitations and recording difficulties. The unique characteristics of dysarthric speech further complicate the annotation process, requiring specialized expertise. Consequently, few open-source dysarthric datasets exist, primarily at the word or command level \cite{uaspeech,easycall}. While some sentence-level dysarthric datasets are available \cite{Torgo,CDSD_arxiv}, their quantity remains limited. To address this data scarcity, dysarthric data augmentation (DDA) is essential. Current sentence-level DDA methods employ either Zero-Shot (simulating general dysarthric characteristics without using target speaker speech data) or All-Test-Data \cite{baseline32022,baseline42024} (utilizing all available test data from the target speaker to train speech generative models and then using these models to synthesize training data) settings. The All-Test-Data setting exhibits the best performance but represents an idealized scenario. Our work focuses on the more practical Zero-Shot and One-Shot settings with speech generative models.

Data augmentation in ASR, which typically involves synthesizing speech from texts in the training set, encounters challenges when applied to DSR. The limited size of dysarthric datasets often results in content mismatches between the training and target domains. Additionally, pronunciation variations between unseen target speakers and those in the training set exacerbate this mismatch. Consequently, synthetic data generated from training set texts using speech generative models trained on either typical speech or existing dysarthric speakers may not accurately capture the characteristics of the target domain. Rosenberg et al. \cite{textcover1} established a performance benchmark by synthesizing speech from test set texts, suggesting that while this data augmentation setting may be impractical for general ASR, it holds value for specific domain applications. We consider sentence-level DSR, a newly developed and challenging task, to be one such application. Chen et al. \cite{textcover2} emphasized the necessity for the texts of synthesized speech to closely align with the target domain to enhance ASR performance. Yang et al. \cite{textcover4} highlighted that text content is more crucial than speaking style in speaker adaptation, positing that performance improvements arise from synthesized speech standardizing the ASR model's language understanding. Based on these insights, we propose a text-coverage strategy (see Fig. \ref{fig:overview}), evaluating its effectiveness in Zero-Shot and One-Shot DDA settings. This strategy utilizes speech generative models to synthesize speech data directly from pre-defined texts. Although limiting the target domain content may not be ideal for general ASR, it offers advantages for applications such as dysarthria rehabilitation or scenarios where dysarthric speakers use common sentences for communication, particularly given the prior focus of DSR on word recognition. Considering the limited dysarthric data and the prosody variability of dysarthric speech, we selected VITS \cite{VITS} for Zero-Shot and Fish-Speech \cite{fish-speech-v1} for One-Shot as our DDA models (see Section \ref{sec:pm} for selection analysis). To our knowledge, this is the first application of these models in the context of DDA.

We evaluated our text-coverage strategy on two sentence-level dysarthric speech datasets: the English Torgo dataset \cite{Torgo} and the Chinese CDSD dataset \cite{CDSD_arxiv}. On the Torgo dataset, the Zero-Shot text-coverage strategy achieved a 3.29\% absolute reduction in average WER, while the One-Shot text-coverage strategy resulted in a 5.98\% absolute WER reduction. For the CDSD dataset, we implemented a stepwise filtering mechanism, ceasing further DDA for speakers with high intelligibility (CER \textless 25\%; \cite{CDSD_arxiv} test set CER: 26.46\%). Finally, 20 out of 24 speakers to achieve a CER below 25\%. By utilizing \textbf{a maximum of one sample} per speaker, the average final CER was 19.525\% (calculated by averaging the boldfaced values in Table \ref{tab:cdsd CER case1-2-3}).

Our text-coverage strategy leverages prior knowledge of the target domain's content and incorporates dysarthric acoustic knowledge from the speech generative model. Related work in text-only domain adaptation \cite{textonly32017,textonly22018} typically employs language models (LMs) trained on target domain text data to improve ASR performance via shallow fusion of ASR and LM outputs during inference. This approach enhances the fluency of predicted text and adapts to the target domain's linguistic characteristics. We compared this approach with ours on the CDSD dataset and found that while it can improve performance, our method yields more significant gains.

The main \textbf{contributions} of this work are as follows:
\begin{itemize}
    \item We systematically analyzed existing DDA methods, highlighting their advantages and limitations.
    \item We present a novel and efficient text-coverage strategy that enables the first sentence-level zero-shot and one-shot DDA based on speech generative models.
    \item We evaluated the effectiveness of our proposed DDA strategy and selected models using the Torgo and CDSD dysarthric datasets.
\end{itemize}

\section{Previous methods}
\label{sec:rw}
\noindent\textbf{Pure signal processing approaches} for dysarthric data augmentation (DDA) encompass implementations such as: (1) modifying speaking rate using PSOLA and pitch using linear transformation \cite{dda20182}, (2) manipulating the log-Mel spectrogram through time warping, frequency and time masking \cite{dda20212}, (3) perturbing vocal tract length and adding reverberation \cite{dda20221}, and (4) applying various spectral-temporal transformations \cite{dda202110}. While these methods simulate general characteristics of dysarthric speech, providing interpretability, their limited prosodic enhancement restricts the generation of authentic, diverse speech, hindering optimal performance.

\noindent\textbf{GANs} Given the significant pronunciation variability among dysarthric speakers, each speaker's pronunciation pattern can be considered a unique distribution. This characteristic has led to the use of various GANs for DDA, including DCGAN \cite{dda20233}, CycleGAN \cite{dda20215}, VAE-GAN \cite{dda20232}, and Spectral basis GAN \cite{dda20241}. While GAN-based DDA excels at fitting data distributions, it requires substantial target speaker data; insufficient data can cause model collapse and limit output diversity.

\noindent\textbf{Reinforcement learning} Jin et al. \cite{dda20242} proposed a dynamic DDA method that utilizes reinforcement learning to optimize SpecAugment parameters during DSR training. While this approach demonstrates good performance through a speech chain, it is dependent on existing data and necessitates sufficient training data, along with further exploration of diverse settings.

\noindent\textbf{Reusing speech generative models} Recent DDA research has adapted Voice Conversion (VC) and Text-to-Speech (TTS) models, originally designed for typical speech, by training or fine-tuning them using dysarthric data. Models that have been adopted include: (1) attention-based VC \cite{dda20213}, (2) sparse-structured spectral envelope transformation-based VC \cite{dda20217}, (3) autoencoder-based VC \cite{dda20218}, (4) diffusion-based VC \cite{dda20236} and TTS \cite{baseline42024}. Beyond directly reusing models, Soleymanpour et al. \cite{baseline32022} modified a multi-speaker TTS system by incorporating a dysarthria severity level coefficient and a pause insertion model, enabling the synthesis of dysarthric speech across varying severity levels. While these methods show promise, Hu et al. \cite{dda20231} demonstrated that some TTS models struggle to generate natural-sounding dysarthric speech, requiring larger datasets or longer training times. Furthermore, with the exception of \cite{baseline32022,baseline42024}, these works primarily focus on enhancing word recognition, and this approach has not fully kept pace with the progress of typical speech TTS models, nor has it explored Voice Cloning technology.

\begin{figure}[t]
\centerline{\includegraphics[width=10cm]{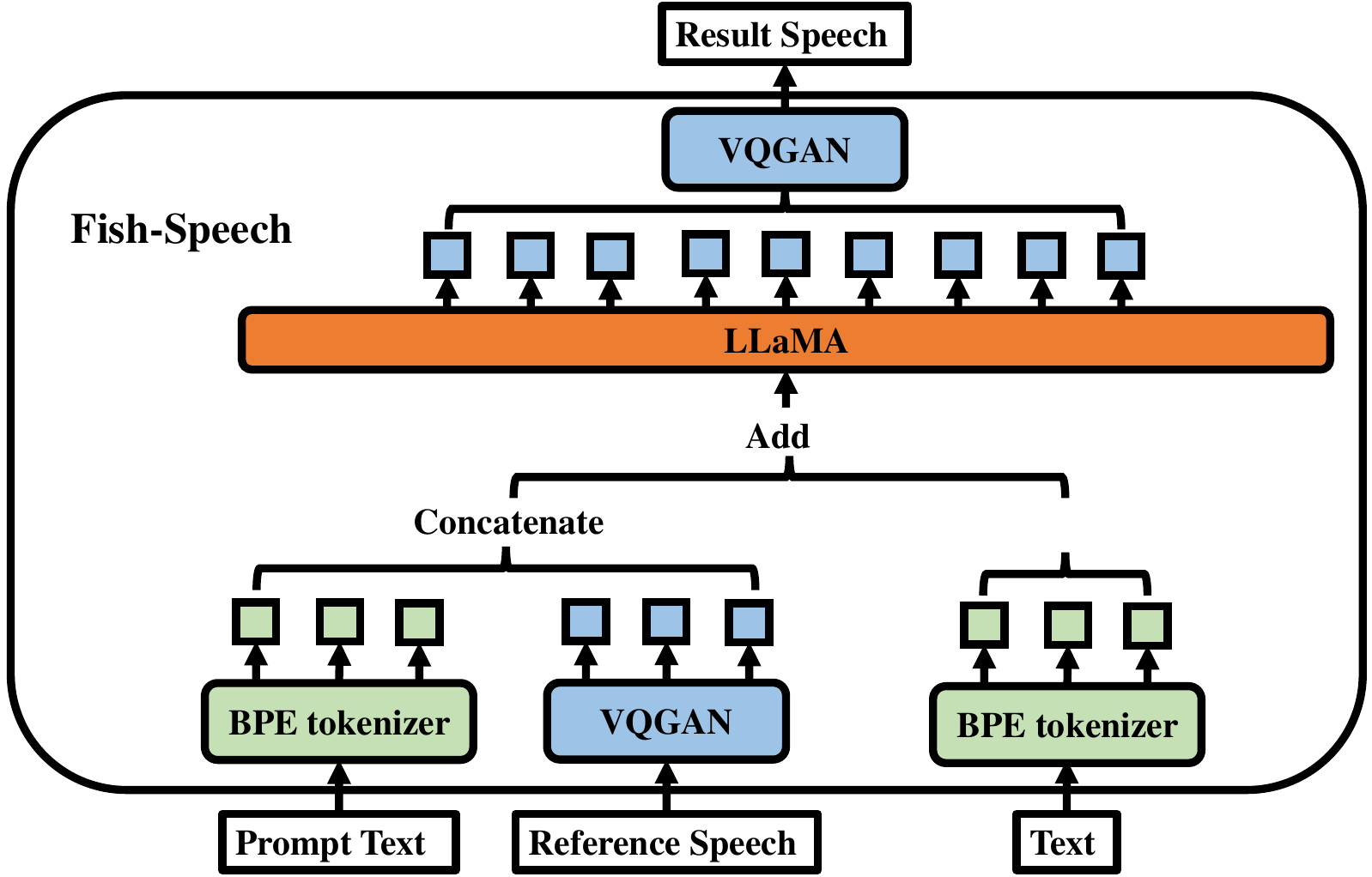}}
\caption{Fish-Speech Inference Framework.`Prompt Text' corresponds to `Reference Speech' and `Text' is the target texts for synthesis. BPE: byte pair encoding.}
\label{fig:f2}
\end{figure}
\renewcommand{\thefootnote}{\arabic{footnote}} 
\section{Strategy design and model selection}
\label{sec:pm}
This section details the rationale for our dysarthric data augmentation (DDA) strategy design and DDA models selection.

\noindent\textbf{Text-coverage strategy} To enhance the performance of Dysarthria Speech Recognition (DSR) models for unseen dysarthric speakers, improving the generalization ability of the DSR model or adapting the model is crucial. However, constructing a speaker-independent and text-independent DSR model requires large and diverse training data. To our knowledge, the number of speakers and texts covered by the widely-used Torgo and CDSD datasets remains limited. Constructing a speaker-dependent DSR model requires a large amount of comprehensive speech data from the target speakers to capture their unique pronunciation patterns and prosodic variations, but collecting and annotating dysarthric data are very difficult. Building on the findings discussed in \cite{textcover1,textcover2,textcover4} in Section \ref{sec:intro}, we propose the text-coverage strategy (see Fig. \ref{fig:overview}). This strategy involves synthesizing speech data based on the pre-defined texts content of the target scenario, allowing the DSR model to adapt proactively and enhance its performance.

\noindent\textbf{VITS} We chose the end-to-end TTS model VITS because it eliminates the need for separate vocoder adaptation, a process requiring extensive data. This is particularly advantageous given the limited availability of dysarthric speech data, as the vocoder significantly impacts generated speech quality and accuracy. Furthermore, VITS's architecture, incorporating GANs, can effectively learn the pronunciation distribution of dysarthric speakers through adversarial training. Moreover, as a conditional variational autoencoder, VITS introduces variability through latent variables and a random duration predictor, generating synthesized speech with diverse prosody for the same input text. This matches the prosodic variations in dysarthric speech and improves DSR model generalization.

\noindent\textbf{Fish-Speech} We selected Fish-Speech, a free and open-source voice cloning model. Its inference framework is illustrated in Figure \ref{fig:f2}, which we created based on the code from \cite{fish-speech-v1}\footnote{The code used is from August 2024.}. Our decision was influenced by several factors. First, Fish-Speech's UTF-8 text encoding and byte pair encoding tokenizer bypasses the phoneme system, eliminating the need for language-specific preprocessing and training, an advantage for our experiments involving two languages. Second, its architecture, a recently evolving neural codec-based architecture \cite{tokenssurvey}, can efficiently generate speech with diverse and realistic prosody. Finally, preliminary testing on the Torgo dataset indicated that its cloning performance for dysarthric speech, evaluated through sampling and manual assessment, is acceptable and could be further improved by fine-tuning LLaMA \cite{LLaMA} of Fish-Speech.

\begin{table*}[t] 
    \caption{WER (\%) on Torgo. `M', `L', and `VL' denote moderate, low, and very low severity, respectively. `FT' indicates fine-tuning. `AVG' is the speaker-level average WER. See Section \ref{sec:expup} for definitions of \textbf{V}, \textbf{F1}, and \textbf{F2}.}
    \label{tab:Torgo wer} 
    \centering
    \begin{tabular}{l | l | c c c | c}
        \toprule
\textbf{Settings} & & \textbf{Severity} & \textbf{Level} & & \textbf{AVG} \\
& & \textbf{M}& \textbf{L} & \textbf{VL} & \\
        \midrule
        & \cite{baseline12016} & 54.46 & 33.89 & 10.35 & 40.86 \\
        & \cite{baseline22017} & 48.40 & 31.07 & 8.66 & 36.30 \\
    \textbf{Without} & \cite{baseline32022} & 56.34 & 46.60 & 13.85 & 44.50 \\
    \textbf{DDA} & \cite{baseline42024} & 85.35 & 28.54 & \textbf{3.44} & 57.77\\
        & LOSO & \textbf{39.56} & \textbf{24.60} & 6.30 & \textbf{29.38} \\
        \midrule
    \textbf{Zero-Shot}
    & \textbf{V} FT LOSO & 34.38 & 24.40 & 6.20 & 26.09 ($\Delta$ \textbf{3.29\%})\\
        \midrule
   \textbf{One-Shot}
    & \textbf{F1} FT LOSO & 33.36 & 21.50 & 5.85 & 25.00 \\
        & \textbf{F2} FT LOSO & \textbf{31.06} & \textbf{20.90} & \textbf{5.50} & \textbf{23.40} ($\Delta$ \textbf{5.98\%})\\
        \midrule
    \textbf{All-Test-Data}
    & \cite{baseline32022} & 50.14 & 36.80 & 12.60 & 39.09 \\
        & \cite{baseline42024} & 26.37 & 26.14 & 3.82 & 20.70 \\
        \bottomrule
    \end{tabular}
\end{table*}
\section{Experiments}
\label{sec:exp}
\subsection{Dataset}
\noindent\textbf{Torgo dataset} The Torgo dataset comprises English speech from 8 dysarthric speakers and 7 control speakers. After filtering out samples with transcription errors and unrestricted content, a total of 16,582 data segments remained, with an approximate duration of 13.67 hours. Using the data from the 8 dysarthric speakers, we fine-tuned the pre-trained model using the leave-one-speaker-out (LOSO) method to construct the LOSO model. For the DDA models, all data except that of the target dysarthric speaker were used during training or fine-tuning, including data from the control speakers.

\noindent\textbf{CDSD dataset} We utilize the initial release of the CDSD dataset, comprising 29,466 Chinese recordings from 24 dysarthric speakers. Using the data from these 24 speakers, we also fine-tune the pre-trained model with the LOSO method to construct a LOSO model for DSR. For the DDA models, we incorporate 12,885 recordings (14.38 hours) from 31 dysarthric speakers with intelligibility annotations from the AISHELL-6B\footnote{\url{https://www.aishelltech.com/AISHELL_6B}} dataset during training or fine-tuning

\subsection{Experimental settings} 
\label{sec:expup}
\noindent\textbf{DDA models} We implemented the VITS model using the open-source framework\footnote{\url{https://github.com/jaywalnut310/VITS}}, which has 39.68M parameters. We followed the original VITS codebase for English text processing, while performing word segmentation and Pinyin parsing for Chinese text. Hyperparameters were adapted from the vctk\_base.json: for Torgo, batch size 16, n\_speakers 14, 64k training steps; for AISHELL-6B, batch size 64, n\_speakers 31, 36k training steps. We used the pre-trained Fish-Speech model\footnote{\url{huggingface.co/fishaudio/fish-speech-1.2-sft}} (532.07M parameters, 150k hours of pre-training data: 50k hours each of English, Chinese, and Japanese, and 1.5k hours of mixed data for supervised fine-tuning). Pre-trained Fish-Speech produced good timbre similarity but poor prosody. Since VQGAN \cite{VQGAN} in Fish-Speech affects timbre and LLaMA affects prosody, we fine-tuned the LLaMA component using LoRA \cite{LoRA} with a learning rate of 1e-10 for 2000 steps. 

\noindent\textbf{DSR models} We trained models using the ESPnet framework\footnote{\url{https://github.com/espnet/espnet}}. Training was set for a maximum of 200 epochs, with early stopping based on validation loss. For Torgo, an ASR model (70.47M parameters) pre-trained on the 960-hour LibriSpeech dataset\footnote{\url{https://www.openslr.org/12/}} was fine-tuned using the LOSO approach. We synthesized speech for the test set texts using VITS, pre-trained Fish-Speech, and fine-tuned Fish-Speech, creating settings \textbf{V}, \textbf{F1}, and \textbf{F2}. Synthesized data was used to further fine-tune the LOSO models. For CDSD, an ASR model (116.91M parameters) pre-trained on the 10k-hour WenetSpeech dataset \cite{WenetSpeech} was fine-tuned with LOSO, followed by additional fine-tuning using the DDA data. To mitigate potential negative impacts of DDA on high-intelligibility speakers (CER \textless 25\%), we implemented a stepwise filtering mechanism. This involved progressing through LOSO models with Zero-Shot (\textbf{V}), One-Shot (\textbf{F2}), and All-Test-Data (\textbf{F3}) settings. Note that \textbf{F3} is identical to \textbf{F2} except that it includes all test data from the target speaker during the Fish-Speech fine-tuning stage. When a model's CER on a target speaker's speech is less than 25\%, the LOSO model for that speaker is excluded from further DDA testing.

\noindent\textbf{Text-only domain adaptation} uses shallow fusion of ASR and language model (LM) predictions during inference:
\begin{equation}
\hat{y} = \underset{y}{\operatorname{argmax}} \log p(y|x) + \lambda \log p_{LM}(y), \label{eq1}
\end{equation}
where $x$ represents the speech features, $y$ is the label sequence, $p(y|x)$ is the ASR model probability, $p_{LM}(y)$ is the LM probability, and $\lambda$ weights the LM's contribution. Due to the limited amount of sentence data in Torgo (25.5\%) and space constraints, we only evaluated text-only domain adaptation on CDSD to assess the impact of text adaptation. We trained ESPnet LMs (54.06M parameters) for 100 epochs using the training and test set texts (representing source and target domains) in the LOSO setup. We explored the effects of $\lambda$ values of 0.3, 0.6, and 0.8.

\subsection{Results and discussions}
\label{sec: result}
The experimental results on the Torgo dataset are presented in Table \ref{tab:Torgo wer}. Espana Bonnet et al. \cite{baseline12016} compared the performance of various DNN architectures on Torgo, but none of the models achieved optimal results for all speakers. For our comparison, we selected the best performance for each speaker from \cite{baseline12016}. Joy et al. \cite{baseline22017} improved DSR performance by adjusting DNN architecture parameters. Soleymanpour et al. \cite{baseline32022} employed a DNN-HMM architecture for DSR, while Leung et al. \cite{baseline42024} utilized the large speech model Whisper \cite{whisper} for DSR. We include the results from \cite{baseline32022,baseline42024} both before and after DDA. DDA methods used in \cite{baseline32022,baseline42024} are described in Section \ref{sec:rw}. Our LOSO models demonstrate strong performance without DDA. The Zero-Shot setting, which employs the DDA method that combines the VITS model with the text-coverage strategy, achieves a significant 3.29\% absolute decrease in average WER. Further improvements are observed in the One-Shot setting, where Fish-Speech utilizes a single speech sample from the target dysarthric speaker to generate data that is more aligned with the target domain, leading to enhanced performance across all severity levels. Compared to the results of \cite{baseline32022,baseline42024} enhanced under the All-Test-Data setting, our One-Shot results significantly outperform \cite{baseline32022} (p-value \textless 0.05) and achieve comparable performance to \cite{baseline42024}. Notably, we achieve these results using only one sample from the unseen target speaker, indicating that our method of leveraging Fish-Speech to learn the speaking style while aligning with the target text domain is an effective DDA approach.

For the CDSD dataset, lacking prior work using the LOSO setup for comparison, we first validated our model settings by building a DSR model. This was achieved by fine-tuning the pre-trained model using the data split described in \cite{CDSD_arxiv}. Our model achieved a CER of 20.4\% on the test split, a 6.06\% absolute reduction compared to the results reported in \cite{CDSD_arxiv}. As shown in Table \ref{tab:cdsd CER case1-2-3}, our LOSO models reduced the CER for 13 speakers to below 21\%. The effectiveness of our DDA methods is evident from Table \ref{tab:cdsd CER case1-2-3}: (1) The Zero-Shot setting further reduced the CER of 4 additional speakers to below 18\% (p \textless 0.05), demonstrating the benefit of combining a multi-speaker dysarthric TTS model with our text-coverage strategy. (2) The One-Shot setting, which combines voice cloning and the text-coverage strategy, further improved DSR performance for the remaining 7 speakers (p \textless 0.05). Overall, we reduced the CER to below 25\% for 20 speakers. However, speakers 06 and 18 (both children) and speakers 11 and 21 (both elderly men) presented greater challenges due to higher severity levels of dysarthria. This highlights the need for exploring new solutions specifically tailored to these more severely affected speakers.

We applied text-only domain adaptation to enhance LOSO models on the CDSD dataset (results in Table \ref{tab:cdsd CER case1-2-3}). By tuning the LM weight, $\lambda$, we consistently achieved performance improvements over the baseline LOSO models. Although the ESPnet framework defaults to a $\lambda$ of 0.3, increasing $\lambda$ generally improved performance for most speakers, highlighting the difficulty LOSO models face with unseen dysarthric speech. Furthermore, for most speakers, LMs trained on the training set texts (source domain) outperformed those trained on test set texts (target domain), likely due to the limited size of the test sets in the LOSO setup. However, our proposed method, which leverages only test set texts and acoustic information from the DDA model, yielded significantly greater improvements. We attribute this to our method's focus on directly improving the LOSO model itself. 

\setlength{\tabcolsep}{0.45mm}{
\begin{table*}[t] 
    \caption{CER (\%) on CDSD. `LM*' denotes enhancing the LOSO models via text-only domain adaptation.}
    \label{tab:cdsd CER case1-2-3} 
    \centering
    \begin{tabular}{l | c c c c c c c c c c c c c}
        \toprule
        \textbf{Speaker ID} & \textbf{04} & \textbf{05} & \textbf{07} & \textbf{09} & \textbf{12} & \textbf{13} & \textbf{14} & \textbf{15} & \textbf{19} & \textbf{20} & \textbf{23} & \textbf{25} & \textbf{26} \\
        \midrule
        Pre-train ASR & 38.8 & 34.9 & 33.3 & 63.2 & 21.5 & 7.0 & 22.1 & 57.3 & 19.0 & 46.3 & 47.5 & 54.5 & 58.2\\
        LOSO models & \textbf{3.0} & \textbf{7.1} & \textbf{7.7} & \textbf{18.8} & \textbf{2.8} & \textbf{1.5} & \textbf{3.0} & \textbf{15.3} & \textbf{14.5} & \textbf{11.8} & \textbf{9.6} & \textbf{18.2} & \textbf{20.9} \\
    \toprule
    \textbf{Speaker ID} & \textbf{01} & \textbf{02} & \textbf{10} & \textbf{16} & \textbf{03} & \textbf{08} & \textbf{17} & \textbf{06} & \textbf{11} & \textbf{18} & \textbf{21} \\   
        \midrule
        Pre-train ASR & 30.3 & 73.8 & 76.6 & 54.1 & 63.8 & 67.1 & 48.1 & 96.0 & 73.8 & 96.0 & 59.8\\
        LOSO models & 26.7 & 40.3 & 43.0 & 36.5 & 40.1 & 49.3 & 37.5 & 87.1 & 56.8 & 94.6 & 43.5 \\
        \textbf{Zero-Shot} & \textbf{8.7} & \textbf{13.3} & \textbf{17.6} & \textbf{14.6} 
        & 29.4 & 28.1 & 29.1 & 71.3 & 40.9 & 92.7 & 36.0\\
        \textbf{One-Shot} & --- & --- & --- & --- 
        & \textbf{17.8} & \textbf{22.2} & \textbf{15.5} & \textbf{63.5} & \textbf{38.2} & \textbf{90.9} & \textbf{32.1} \\
        \textbf{All-Test-Data} & --- & --- & --- & --- 
        & --- & --- & --- & 58.0 & 36.4 & 90.7 & 31.1 \\
        $LM_{source}$ $\lambda=0.3$ & 20.4 & 32.5 & 32.1 & 38.6 & 36.5 & 36.4 & 40.7 & 86.9 & 48.5 & 94.4 & 37.5\\
        $LM_{source}$ $\lambda=0.6$ & 19.1 & 28.3 & 25.5 & 42.8 & 36.5 & 27.9 & 45.8 & 83.9 & 45.9 & 94.2 & 35.0 \\
        $LM_{source}$ $\lambda=0.8$ & 19.8 & 28.7 & 24.1 & 47.5 & 37.6 & 26.0 & 49.7 & 83.0 & 46.8 & 94.2 & 35.7\\
        $LM_{target}$ $\lambda=0.3$ & 24.8 & 37.3 & 40.6 & 34.7 & 39.1 & 45.0 & 36.4 & 87.5 & 52.7 & 93.4 & 41.3 \\
        $LM_{target}$ $\lambda=0.6$ & 23.9 & 35.2 & 38.4 & 33.9 & 38.8 & 42.6 & 37.0 & 84.5 & 51.6 & 92.2 & 40.8\\
        $LM_{target}$ $\lambda=0.8$ & 24.1 & 35.4 & 38.7 & 34.1 & 39.4 & 42.5 & 38.5 & 83.9 & 52.8 & 92.1 & 41.4\\
    \bottomrule
    \end{tabular}
\end{table*}
}

\section{Conclusions}
\label{sec:conclusions}
We introduce a novel text-coverage strategy to improve sentence-level dysarthric speech recognition for unseen speakers in constrained scenarios.
This strategy, combined with VITS and Fish-Speech, enables the first sentence-level zero-/one-shot dysarthric data augmentation using speech generative models. We have validated our strategy on the English dysarthric dataset Torgo and the Chinese dysarthric dataset CDSD.

\subsubsection{Acknowledgements} 
This work has been supported in part by  NSF China (Grant No.62271270).

%
%
%
%

\bibliographystyle{IEEEbib}
\bibliography{refs}


\end{document}